\documentclass[letterpaper]{jpconf}

\begin{document}
\title{A non-linear irreversible thermodynamic perspective on organic pigment proliferation and biological evolution}
\author{K.~Michaelian}
\address{Instituto de F\'{\i}sica, Universidad Nacional Aut\'{o}noma de
M\'{e}xico, Cto.~de la Investigaci\'{o}n Cient\'{\i}fica,
Cuidad Universitaria, C.P.~04510, Mexico}
\ead{karo@fisica.unam.mx}


\begin{abstract}

The most important thermodynamic work performed by life today is the dissipation of the solar photon flux into heat through organic pigments in water. From this thermodynamic perspective, biological evolution is thus just the dispersal of organic pigments and water throughout Earth's surface, while adjusting the gases of Earth's atmosphere to allow the most intense part of the solar
spectrum to penetrate the atmosphere and reach the surface to be
intercepted by these pigments. The covalent bonding of atoms in organic pigments
provides excited levels compatible with the energies of these photons. Internal conversion through vibrational relaxation to the ground
state of these excited molecules when in water leads to rapid dissipation of
the solar photons into heat, and this is the major source of entropy
production on Earth. A non-linear irreversible thermodynamic analysis shows that the proliferation of organic pigments on Earth is a direct consequence of the pigments catalytic properties in dissipating the solar photon flux. A small part of the energy of the photon goes into the production of more organic pigments and supporting biomass, while most of the energy is dissipated and channeled into the hydrological cycle through the latent heat of
vaporization of surface water. By dissipating the surface to
atmosphere temperature gradient, the hydrological cycle further increases
the entropy production of Earth. This thermodynamic perspective of solar photon dissipation by life has implications to the possibility of finding extra-terrestrial life in our solar system and the Universe.
\end{abstract}

\section{Introduction}

There are two kinds of stable (time persistent) macroscopic organizations
found in nature. The first are called ``equilibrium structures'' and result from the
minimization of free energy subject to the conservation laws of
physics arising from the symmetries in nature. These organizations have little interaction with their environment and are in states of near maximum entropy with zero entropy production. Examples are
the crystalline structure of condensed material or the planetary orbits of our solar
system. 

The second kind of organizations are called ``dissipative structures'' and are asymptotically stable (attractive) and are much richer in
complexity and variety, have important interaction with their environment, and are the result of the dissipation of a
thermodynamic potential, themselves not subject to the conservation laws although their overall interaction with their environment is. The general non-linear nature of these systems means that they can be in any of a multitude of macroscopically different locally stable states dependent on their particular history, and tend to evolve towards states of greater entropy production. Examples are ecosystems, social systems, climate systems, and global elemental cycle systems including the atmosphere and ocean systems.

Building on the non-equilibrium thermodynamic foundation established by Th{\'e}ophile de Donder (1936) and  Lars Onsager (1931) ,
Ilya Prigogine (1967) and co-workers have developed a comprehensive framework for dealing
with this second kind of organization. Their framework, known as {\it Classical Irreversible Thermodynamics} (CIT), employs the same thermodynamic variables as does equilibrium thermodynamics but allows the variables to vary locally in both space and time. The whole domain of validity of classical irreversible thermodynamics is thus restricted to situations in which the
existence of a local equilibrium can be supposed, i.e. equilibrium attained within a small, but still macroscopic region (of the order of 10$^{23}$particles) of the
system under study. This restriction is required in order to retain the usual meanings of the thermodynamic variables and thus the validity of, for example, the Gibb's equation relating them. It turns out, however, that this domain is surprisingly large and the framework of classical irreversible thermodynamics can be used to treat most macroscopic dissipative phenomena which are common to our everyday experience. 

Employing the CIT formalism, this paper aims to demonstrate that the spread organic pigments and water over the Earth's surface arises as a result of the thermodynamic imperative of dissipating high energy solar photons into numerous low energy ones. The affinity for this process will be shown to be related to a photon chemical potential corresponding to the solar photon spectrum at the Earth's surface, and that corresponding to the emitted and reflected Earth spectrum at the black-body temperature of  the cloud tops which radiate into space. It is argued that co-evolution of the biotic with the abiotic on Earth has promoted the spread of organic pigments and water over Earth's surface, allowing Earth, in interaction with its solar environment, to evolve towards states of greater global entropy production. 

What is thermodynamic reason for pigment proliferation related to dissipation? How is present life related to past life from the perspective of dissipation? What is the importance of water to dissipation? Should dissipating life forms be common throughout the universe? Is life as we know it on Earth inevitable? These are some of the questions to be addressed in this paper. First, however, we discuss the relation between life, water, and the physical and chemical characteristics of the biosphere and its evolution, much of which has been uncovered while assessing the theory of Gaia by James Lovelock (1988) and collaborators. In section \ref{sec:photon} the physics of photon absorption and dissipation by
organic molecules in water is discussed. In section \ref{sec:autocatalysis} we employ the formalism of non-linear irreversible thermodynamics to show how organic pigments are the products of an auto-catalytic cycle which augments the solar photon dissipation, leading to the expectation of the proliferation of these organic pigments and water throughout Earth's surface, and even over other solar system bodies. In section \ref{sec:entropyprod} we show how Earth dissipates about 70\% more than either of its neighboring planets and that most of the dissipation occurs on Earth's surface, building a case for the importance of life in the global dissipation process. Section \ref{sec:originlife} demonstrates, from within this thermodynamic perspective, how present life could be associated with life at its beginnings in the Archean and reviews a recent theory on the dissipative origin of life (Michealian, 2009b, 2011b). Section \ref{sec:lifeuniverse} discusses the possibility of finding similar forms of dissipating life throughout the universe. Conclusions are presented in section \ref{sec:conclusions}.

\subsection{Life, water and the biosphere}

In 2005, NASA crash landed a deep impactor onto the surface of comet Tempel 1 in order to study the gas and dust released using instruments on board the mother ship (A'Hearn et al., 2005). Over 200 organic molecules were found in the dust and gas released. Some precursors to the aminos of amino acids, principally hydrogen cyanide and methyl cyanide, were clearly identified (Moulton, 2005). Other more complex organics were also found but the resolution of the instruments was insufficient to unequivocally identity many of these. The surface of the comet was found to be very dark, having an albedo of only 0.04 (visible wavelength), due, in no small part, to the absorption of light by these organic molecules.The most common amino acid of life, glycine, was found in samples returned to Earth by NASA's Stardust mission which flew by Comet Wild 2 in 2004. Nucleobases, the aromatic ring components of nucleic acids RNA and DNA, have been found in meteorites (Callahan et al., 2011). 

Comets consist mostly of ice (50\% for Tempel 1) and dust and it is probable that during the Archean their delivery contributed a substantial portion of Earth's water. Comets and meteorites could, therefore, have been the providers of both the vital elements of the primordial soup; organic molecules and water (Hoyle and Wickramasinghe, 1978). Alternatively, organic molecules could also have been created in the Earth's somewhat reducing early atmosphere subjected to intense UV light and lightening (Miller and Urey, 1959).

Life has kept the amount of water on Earth's surface relatively constant since its beginnings 3.8 billion years ago. There are physical mechanisms active on all planets that disassociate water into its oxygen and hydrogen components; for example, through lightening or UV light. Hydrogen, being a light element can be carried along by the solar wind and lost into space. Earth's magnetic field helps to shield the Earth from the solar wind and so less hydrogen has been lost as compared to Venus for example, which has been very dry since loosing its magnetic shield about 2 billion years ago. However, perhaps more important in retaining water on Earth is photosynthetic life's ability to release oxygen from carbon dioxide, thereby providing the potential for its recombination with hydrogen. For example, aerobic chemoautotrophic bacteria  oxidize hydrogen sulfide to produce elemental sulfur and water as waste products (Lovelock, 1988). Also, the oxygen released by life can undergo photochemical reactions in the upper atmosphere and be converted into ozone, which then protects water and methane in the lower atmosphere from UV disassociation. 

By controlling the amount of greenhouse gases in the atmosphere, principally water, carbon dioxide, and methane, life has controlled the temperature of Earth's surface, keeping it within a narrow range for water to be present in its liquid phase. This control is surprising considering that, according to the standard solar model, the integrated solar output has probably increased  by as much as 30\% since the beginnings of life on Earth (Sagan and Chyba, 1997). By studying Sun like proxy stars of different ages, the Sun in Time Program (Ribas, 2005) has determined that the amount of far ultraviolet light reaching Earth's atmosphere at the beginning of life could have been up to 30 times greater than today, extreme ultraviolet and soft x-ray perhaps 100 times greater, and x-ray perhaps 2000 times greater (Ribas, 2005). The reason for this has to do with the higher rotation rates of younger stars which gives them turbulent eddies that mix different layers, bringing the hotter inner layers to the surface.

Utilizing the vapor pressure deficit caused by evaporation at the leaves, the roots of plants draw up water through the inter-cellular cavities and into the leaves from where it enters the water cycle. Plants on land increase the amount of water in the water cycle by a factor of almost two (Kleidon, 2008). Almost all of the free energy arriving in the solar photon flux at the leaves is used in evaporation of water, while less than 0.2\% is used in photosynthesis (Gates, 1980), i.e. converted into covalent chemical bonds of organic material. Over water, cyanobacteria play the same role as plants over land, heating the upper surface of the water body and thereby increasing evaporation rates (Jones et al., 2005). The water cycle is a second dissipative process, coupled to photon dissipation by organic pigments, which dissipates the temperature gradient between the warm surface of the Earth and the cold upper atmosphere. The water cycle also helps move water over land masses where it can be used by organic pigments to dissipate solar photons.

Life has gradually adjusted the gases of Earth's atmosphere in such a manner so as to obtain a transparency for the most intense (enthropically speaking) part of the solar spectrum such that it can be intercepted by organic pigments on the surface which converts it into infrared photons that can be readily absorbed by water. Evolution has continually invented new pigments, absorbing ever more completely the solar spectrum (Michaelian, 2011a). 

Cnossen et al. (2007), following Sagan (1973), have shown that, using the best present knowledge for the gases of Earth's Archean atmosphere at the beginning of life on Earth, principally CO$_2$, N$_2$, H, and some CH$_4$, the atmosphere would have had a window of transparency of between 200 and 300 nm (although perhaps reduced to between approximately 240 nm and 290 nm taking into account the aldehydes formed by UV photochemical reactions on these gases (Sagan, 1973)). It is then probable that life on the surface obtained its thermodynamic reason for being by dissipating photons within this wavelength region. DNA and RNA, in fact, absorb very strongly at 260 nm and dissipate the excitation energy rapidly into heat. A theory for the origin of life based on the extraordinary dissipation properties of the nucleic acids in the UV region has been given by Michaelian (2009b, 2011b) and will be reviewed in section \ref{sec:originlife}.

Photons, life, and water thus share an intimate relation, probably since the beginnings of life, based on entropy production. Over land or over water, plants or cyanobacteria dissipate high energy photons into heat. The details of this relation will be reviewed in the following sections.

\subsection{Organic pigments and photon dissipation}
\label{sec:photon}
Organic molecules differ from inorganic molecules in the nature of their
chemical bonding. Atoms in organic molecules are bound by covalent bonds
which are both strong and directional. The directionality of these bonds
allows for an almost infinite set of stable configurations with distinct physical, chemical, and electronic properties. 
Inorganic bonding, on the other hand, is usually of ionic, metallic, or of the van
der Waals type. These central interactions have no
directionality characteristics and therefore configurations of numerous atoms tend to be constrained to compact spherical clusters, with only slight variation of their properties with the size of the cluster.

 Organic pigments absorb strongly over the visible and ultraviolet regions.
They can be classified into three groups according to their wavelength absorption characteristics.  The first group absorbing strongly in the far ultraviolet, (below approximately 300 nm) consists of the aromatic rings where the promotion of an electron from a $\pi$-bonding molecular orbital to a $\pi$-anti-bonding orbital, referred to as a $\pi,\pi$* transition, and has high oscillator strength and a large extinction coefficient (absorptivity). Examples are the nucleic acid bases adenine, guanine, cytosine and thymine, and the amino acids phenylalanine, tryptophan, histidine and tyrosine. The second group consists of the mycosporines and mycosporine-like amino acids (MAAs) characterized by a cyclohexenone or cyclohexenimine chromophore conjugated with the nitrogen substituent of an amino acid and have UV-absorption maxima in the range 310 to 360 nm. The third group absorbing in the visible region are the porphyrins, which are organic rings conjugated by a metallic ion, such as magnesium in chlorophyll, the carotenoids, flavonoides, etc., or the phycobilins found in cyanobacteria. Due to their covalent bonding, organic pigments are thus predilect to form the foundations of a photon dissipative system relevant over a large region of the solar spectrum. 

The first two groups absorbing in the ultraviolet have ultrafast (picosecond) non-radiative de-excitation rates compatible with internal conversion through a conical intersection of the excited state with the ground state (Pecourt et al. 2000, Conde et al., 2000). A conical intersection results when the vibrational states superimposed on the electronic first excited state overlap in energy with the vibrational states superimposed on the electronic ground state because of a slight deformation of the molecule. The de-excitation time through vibrational dissipation are significantly reduced if the molecules are in water since the vibrational energy can be dissipated rapidly through resonances with the high frequency vibrational modes of the water molecule. From this dissipative perspective, it is clearly this fact, more than any other, that has defined life's association with water.

Such rapid dissipation of the excitation energy makes these molecules resistant to destruction by UV light since there simply is not enough time for photochemical reactions to occur. These molecules are thus very robust and efficient entropy producers since they dissipate the strongly absorbed photon energy to heat that can be absorbed by water and are ready to receive a new photon on the order of picoseconds. It is interesting to note that, for example, the non-natural tautomers of the nucleic acid bases have orders of magnitude longer lifetimes and are affected by non-radiationless decay modes including fluorescence and photochemical destruction (Serrano-And\'es and Merch\'an, 2009).

\section{Non-linear Irreversible thermodynamic model}
\subsection{Proliferation of organic pigments due to their catalytic properties in dissipating the solar photon flux}
\label{sec:autocatalysis}

An auto-catalytic reaction is one in which at least one of the products acts as a catalyst for the reaction. Most reactions performed by life today are auto-catalytic. Examples are; protein folding in which the protein FKBP catalyses its own folding (Gadgil and Kulkarni, 2009), or the glycosis cycle which produces ATP from glucose and is auto-catalyzed by the enzyme phosphofuctokinase (PFK). It is generally believed that the first chemical reactions of life must also have been auto-catalytic. Auto-catalytic reactions, like all chemical reactions, arise to dissipate a chemical potential. 

We now consider the production of organic pigments (e.g. nucleic acid bases, mycosporine-like amino acids, porphyrins, etc.) as an auto-catalytic photochemical reaction promoting the dissipation of the solar photon flux. It will be shown that if organic pigments are good catalysts for the irreversible process of photon dissipation, then their concentration will increase on the Earth's surface, over and above that which would be expected under equilibrium conditions. This proliferation of organic pigments over the surface of Earth is basically what has defined biological evolution. The derivation will be similar to that given by Prigogine (1967) for a purely chemical auto-catalytic reaction. However, instead of affinities derived from the chemical potentials dependent on material concentrations of the chemical constituents, we use an affinity derived from the photon chemical potentials which are dependent on the photon gas pressure (Herrmann amd W\"urfel, 2005) and which go as the fourth power of the temperature of the gas for an equilibrium distribution of the photons (black-body spectrum). 

The generalized flows corresponding to these generalized forces defined by the affinities are the flows of energy from one spectrum to another. We assume that the rate of energy conversion will be proportional to the difference of the photon pressures of the different spectra. There are three relevant photon pressures, 1) $P_S$ corresponding to the photon spectrum arriving at Earth, approximately black-body with a temperature equal to that of the surface of the Sun, 2) $P_E$ corresponding to the emitted photon spectrum by Earth, an approximate black-body spectrum with temperature equal to that of Earth's surface, and 3) $P_C$ corresponding to the temperature of the cloud tops at which Earth emits radiation into space.

As a cautionary note, it is emphasized that using a black-body approximation is wrought with error since the radiation arriving at the Earth's surface from the Sun is very directional and only approximates a black-body spectrum. However, the purpose of the calculation is to show qualitatively how the proliferation of organic pigments on Earth can be explained from a non-linear irreversible thermodynamic analysis of the photon dissipation process.

The formalism of classical irreversible thermodynamics (Prigogine, 1967) gives that the entropy production is a product of generalized forces, $X$, times generalized flows, $J$,
\begin{equation}
{\cal P} = {\frac{d_iS }{dt}} = \sum_k J_k X_k \ge 0.
\end{equation}
For chemical reactions, the generalized force is the affinity over the temperature $A/T$, where the affinity $A$ for the reaction is equal to the stoichiometric coefficients of the reactants multiplied by their chemical potentials plus the same of the products. The generalized flow is the rate of the reaction. For our photon dissipation reaction to be considered here, the generalized forces are the photon spectrum affinities determined from their pressures and the generalized flows are the flows of energy from one spectrum to another. 

The time change of the entropy production $d{\cal P}$ can be decomposed into two parts, one related to the
change of forces and the other to the change of flows 
\begin{equation}
d{\cal P} = d_X{\cal P} + d_J{\cal P} = \sum_k J_k dX_k + \sum_k X_kdJ_k.
\end{equation}

In the whole domain of the validity of thermodynamics of irreversible processes, and under constant external constraints, the contribution of the time change of the forces to the entropy production is negative or zero, 
\begin{equation}
d_{X}{\cal P}\le 0.
\end{equation}
This is known as the {\em general evolutionary criterion} and was established by Prigogine (1967). For systems with constant external constraints, the system will eventually come to a steady state in which case (Prigogine, 1967)
\begin{equation}
d_{X}{\cal P}= 0.
\label{eq:genev}
\end{equation}

With this brief background into classical irreversible thermodynamics, consider now the following irreversible processes consisting of the conversion of energy through different photon spectra. First, the flow of energy of the photon spectrum coming from the surface of the Sun $I_S(\lambda)$, which can assumed to be black-body for convenience, $I(T_S)$, with $T_S$=5760 K, is converted, by absorption and dissipation of organic pigments in water, into the emitted photon spectrum of the Earth's surface (also assumed to be black-body), $I(T_E)$ with $T_E$= 287 K, which is then converted into the emitted photon spectrum of the cloud tops $I(T_C)$, with $T_C$= 259 K. We assume that a photochemical reaction takes place as part of the first dissipation process which creates organic pigments of concentration $C$. These pigments themselves act as catalysts for the first dissipation process of $I(T_S) \rightarrow I(T_E)$ making the process auto-catalytic. A schematic diagram for this auto-catalytic process can be given as follows,  
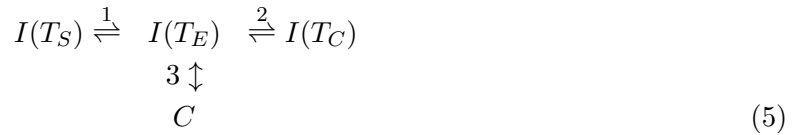
\begin{eqnarray}
I(T_S)\stackrel{1}{\rightleftharpoons } &I(T_E)&\stackrel{2}{\rightleftharpoons }I(T_C) \nonumber \\
&3\updownarrow& \nonumber\\
&C&  \label{eq:reaction}
\end{eqnarray}
Since the system is far from equilibrium, the backward rate constants can be considered as being essentially zero.

The conversion of the energy through the different spectra and the energy of formation of the pigments can be characterized, in a first approximation, by the pressures corresponding to the assumed black-body spectra. We can therefore write the above schematic process alternatively in terms of pressures,
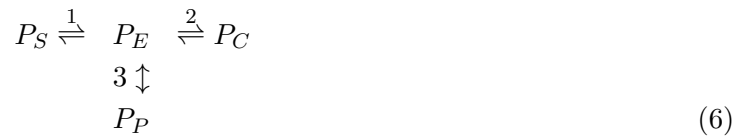
\begin{eqnarray}
P_S\stackrel{1}{\rightleftharpoons } &P_E&\stackrel{2}{\rightleftharpoons }P_C \nonumber \\
&3\updownarrow& \nonumber\\
&P_P&  \label{eq:reaction}
\end{eqnarray}
where $P_P$, the pressure of the organic pigments, is related to their concentration $C$.

In terms of the affinities and flows of the three different processes, we can write the general evolutionary criterion, Eqn. (\ref{eq:genev}), once arriving at the stationary state, in the following form
\begin{equation}
d_{X}{\cal P}=d{\cal P}-d_{J}{\cal P}=d\left(\sum_{\rho=1}^3\frac{A_{\rho }}{T_{\rho }} v_{\rho }\right)-\sum_{\rho=1}^3 \frac{A_{\rho }}{T_{\rho }} dv_{\rho
}= 0
\label{eq:evcrit}
\end{equation}
where $A_1$ is the affinity for the conversion of energy of the solar photon spectrum into energy of the Earth surface photon spectrum, $A_2$ is the affinity for the conversion of energy of the Earth surface photon spectrum to energy of the cloud top photon spectrum, and $A_3$ is the affinity for the photochemical reaction producing the organic pigment. 

For the case of equilibrium photon distributions (black-body spectra), the affinities will go as the logarithm of the ratio of the photon pressures (Herrmann amd W\"urfel, 2005) which are proportional to the temperatures to the fourth power,
\begin{equation}
A_1=kT_E\log{P_S \over P_E}\ \ \ \ \   
A_2=kT_C\log{P_E \over P_C}\ \ \ \ \ 
A_3=kT_P\log{P_E \over P_P}.
\label{eq:forces}
\end{equation}
 
The $v_\rho$ in Eqn. (\ref{eq:evcrit}) are the rates of the corresponding energy conversion (dissipation) processes which, of course, are related to the amount of photon-material interaction which we assume are related to the differences of the photon pressures attributed to the different spectra. A more analytic justification for this comes from the continuity equation for momenta $p$ in continuous media (a kind of Navier-Stokes equation without external forces and without viscosity for photons)
\begin{equation}
{dp \over dt} \propto -{dP \over dx},
\end{equation}
and, since for photons $E=pc$, the energy conversion rate between spectra would go like
\begin{equation}
v={dE \over dt} \propto -{dP \over dx}.
\end{equation} 

Since the organic pigments are assumed to act as catalysts for the conversion of energy from the solar spectrum to energy of the Earth surface spectrum, the rate of the first dissipation process, $I(T_S) \rightarrow I(T_E)$, is multiplied by a factor $(1+\alpha P_P)$ where $\alpha$ represents the effectiveness of the organic pigment as a catalyst for energy conversion (i.e. $\alpha \rightarrow \infty$ for an excellent catalyst, and $\alpha \rightarrow 0$ for a completely ineffective catalyst). Therefore, the rates of conversion, assuming all constants of proportionality equal to one for convenience (again, we are only interested in showing qualitatively the dynamics of pigment proliferation), are given by
 \begin{equation}
v_1= (1+\alpha P_P)(P_S - P_E)\ \ \ \ \   
v_2=P_E - P_C\ \ \ \ \ v_3=P_E-P_P
\label{eq:rates}
\end{equation}
Note the non-linear relation between the forces, Eq. (\ref{eq:forces}), and flows, Eq. (\ref{eq:rates}). Such non-linearity is what gives rise to multiple solutions for the steady state when the system is far from equilibrium.

Using Eq. (\ref{eq:evcrit}) for the steady state together with Eqs. (\ref{eq:rates}) and (\ref{eq:forces}), taking the Boltzmann constant $k=1$ for convenience, and observing that the free forces can be characterized in terms of the two free pressures, $P_E$ and $P_P$ (since $P_S$ and $P_C$ are fixed and given by the Sun surface temperature to the fourth power and cloud top temperature to the fourth power respectively) gives
\begin{eqnarray}
\frac{\partial }{\partial P_E}\left[(1+\alpha P_P)(P_S-P_E)\log\frac{P_S}{P_E} + (P_E-P_C)\log\frac{P_E}{P_C}+(P_E-P_P)\log\frac{P_E}{P_P} \right] &\nonumber\\
+(1+\alpha P_P)\log\frac{P_S}{P_E}-\log\frac{P_E}{P_C}-\log\frac{P_E}{P_P}=0&
\end{eqnarray}
\begin{eqnarray}
\frac{\partial }{\partial P_P}\left[(1+\alpha P_P)(P_S-P_E)\log\frac{P_S}{P_E} + (P_E-P_C)\log\frac{P_E}{P_C}+(P_E-P_P)\log\frac{P_E}{P_P} \right] &\nonumber\\
-\alpha (P_S-P_E)\log\frac{P_S}{P_E}+\log\frac{P_E}{P_P}=0&
\end{eqnarray}
which, after algebra gives for the steady state,
\begin{equation}
v_{1}=v_{2},\ \ \ v_{3}=0,
\end{equation}
and (for one of the solutions of the steady state)
\begin{eqnarray}
P_P=P_E &=&{\frac{1}{2\alpha }}[\alpha P_S-2+[4+4\alpha P_S(1-\gamma )+\alpha
^{2}P_S^2]^{\frac{1}{2}}]  \nonumber \\
&\rightarrow &{\frac{1}{2}}(P_S+P_C)\ \ \ for\ \ \ \alpha \rightarrow 0 
\nonumber \\
&\rightarrow &P_S\ \ \ for\ \ \alpha \rightarrow \infty  \label{eq:concens}
\label{eq:limits}
\end{eqnarray}
with $1-\gamma\equiv P_C/P_S$ ($\gamma$ is, therefore, a measure of the ``distance" from equilibrium of the system). Therefore, since $P_S$ is much greater than $P_C$ (the pressures go as the temperature to the fourth power for black-body spectra) equation (\ref{eq:limits}) indicates that the pressure of the organic pigments $P_P$, or in other words its concentration $C$, has increased due to its catalytic activity in dissipating the solar photon spectrum into the Earth emitted spectrum. 

The entropy production of the energy conversion processes, including catalytic activity of the organic pigment, is given by
\begin{equation}
{\frac{d_{i}S}{dt}}=\sum vA =(P_S-P_E)(1+\alpha P_P)\log {\frac{P_S}{P_E}}+(P_E-P_C)\log {\frac{P_E}{P_C}}+(P_E-P_P)\log {\frac{P_E}{P_P}}.
\end{equation}
Although it will not be demonstrated here, it can also be shown that the entropy production at the stationary state shifts to larger values as a result of the catalytic
activity (see Prigogine, 1967, for the corresponding case of purely chemical reactions). 

These results give a non-linear irreversible thermodynamic explanation for the proliferation of organic pigments over Earth's surface. Pigment concentrations can therefore attain values much greater than that expected in equilibrium, depending on the ratio of $P_S/P_C$. 

Given the above derivation, and now imagining a much more complex system with many coupled irreversible processes operating, it is not difficult to visualize an associated biotic-abiotic co-evolution of Earth's physical and chemical characteristics towards non-equilibrium thermodynamic stationary states with ever greater global entropy production for Earth in its interaction with its solar environment. This thermodynamic explanation of organic pigment proliferation has hitherto been characterized as biological evolution acting through natural selection.

In referring to purely chemical reactions, Prigogine (1967), in fact, noticed that such a result may shed light on the problem of the occurrence of complicated biological molecules in steady state concentrations which are of orders of magnitude larger than the equilibrium concentrations. In his 1967 book ``Thermodynamics of Irreversible Processes" (Prigogine, 1967) Prigogine states ``Thus, for systems sufficiently far from equilibrium, kinetic factors (like catalytic activity) may compensate for thermodynamic improbability and thus lead to an enormous amplification of the steady state concentrations. Note that this is a strictly non-equilibrium effect. Near equilibrium, catalytic action would not be able to shift in an appreciable way the position of the steady state." 

An example of a present day auto-catalytic photochemical reaction tied to solar photon dissipation is that of UV light induced mycosporine production (Sinha et al., 2002). The pigments mycosporine can be considered as catalysts that promote the dissipation of UV light into heat. The biologists see this UV induced mycosporine production as an evolved response of a plant or cyanobacteria to protect itself from its harsh environment. However, it is more likely the production of mycosporine under UV light, or of chlorophyll under visible light, has nothing to do with imaginary ``vital" forces underlying a metaphysical ``will to survive" but rather with non-linear irreversible thermodynamic imperatives founded on the well known and well characterized forces and symmetries of nature.

\section{Results}
\subsection{Global entropy production}
\label{sec:entropyprod}

Planck's formula (Planck, 1913) for the entropy flow of an arbitrary beam of photons is (Wu et al., 2011), 
\begin{eqnarray}
L (\lambda) & = &{n_0 k c \over \lambda^4} \left[\left(1+{\lambda^5I(\lambda) \over n_0 h c^2}\right)\ln\left(1+{\lambda^5I(\lambda) \over n_0 h c^2}\right) - \left({\lambda^5I(\lambda) \over n_0 h c^2}\right)\ln\left({\lambda^5I(\lambda) \over n_0 h c^2}\right)\right],
\label{eq:entropyflux}
\end{eqnarray}
where $n_0=2$ for an unpolarized photon beam and $n_0=1$ for a polarized beam and the units are [J/(m$^3$$\cdot$K$\cdot$ s)]. Using this equation
and approximating the incoming solar spectrum at the top of Earth's atmosphere as a black-body spectrum at temperature of the Sun's surface (5760 K) and the Earth's out going spectrum as a gray-body spectrum at an equivalent temperature of 254.3 K, the total entropy production of Earth can be calculated to be about 1.196 [W/(m$^2 \cdot K$)] averaged over day and night and Earth's entire surface (Michaelian, 2012). Earth's entropy production per square meter, in fact, is found to be roughly 70\% larger than either that of Venus and Mars,(Michaelian, 2012). It is tempting to assign this ``additional" entropy production to the process of life. But, if life is indeed responsible for this additional entropy production, then a significant portion of Earth's total entropy production would have to occur at its surface where the organic pigments are located.

A rough estimate can be made of the relative contribution of the entropy production at the surface of Earth to the total by assuming a heat flow approximation for entropy production;
\begin{equation}
\sigma=Q\left({1\over T_2} - {1\over T_1} \right).
\label{eq:entropyprod}
\end{equation}
Of the solar energy incident on Earth, with a global average of about 238 [W/m$^2$] impinging on Earth's upper atmosphere, about $Q=170$ [W/m$^2$] makes it to the Earth's surface, while the remaining 68 [W/m$^2$] is absorbed in the atmosphere. Using equation (\ref{eq:entropyprod}) with $T_1=5760$K the temperature of the Sun's surface, and $T_2=288$K as the temperature of the Earth's surface, and $T_2=252$K for the temperature of the Earth's atmosphere, one can easily show (Michaelian, 2012) that surface dissipation makes up 71\% of the entropy production and atmospheric dissipation the remaining 29\%. Since organic pigments are densely spread everywhere on Earth's surface where there is water, it follows that life, through organic pigments in water, is probably very important to the entropy production of Earth.

It might be inquired as to what is the function of animals if the main thermodynamic function of life is the dissipation of solar photons through the spread of organic pigments over Earth's surface? In terms of biomass or number, animals are negligible compared with photosynthetic organisms and thus, at first sight, may appear as a mere curiosity. However, their existence appears to be crucial in allowing plants and cyanobacteria to spread over the entire surface of Earth. For example, most of the ocean surface would be as barren as a desert in nutrients, unable to support surface cyanobacteria, if it were not for the mobility of marine animals that spread nutrients over vast distances of ocean surface through mobility, excrement and death. Over land, animals play this same role of the faithful, but unwitting, gardener. Indeed, it has always been argued that animals brought nutrients to the barren land surface perhaps 700 million years ago when life first left the ocean and organic pigments began to proliferate on land (Pisani et al., 2004).  

\subsection{The dissipative origin of life}
\label{sec:originlife}

If the reason for life is the dissipation of solar photons through the spread of organic pigments over the Earth's surface, then it would of course be interesting to look for the primordial organic pigments which could have represented life's initiation. Fortunately, the search appears trivial since the presently presumed first molecules of life, RNA and DNA, both absorb very strongly at 255 nm, just where Sagan (1973) and Cnossen et al. (2007) have predicted a peak in the transparency of Earth's atmosphere during the Archean. As mentioned in section \ref{sec:photon}, the excitation energy due to a photon absorbed on RNA or DNA is dissipated into vibrational energy of the surrounding water very rapidly, making these molecules excellent photon dissipaters in the UV. 

The reason for the proliferation of these molecules under an intense UV flux is thermodynamic and was given in section \ref{sec:autocatalysis}. However, knowing the particular mechanism of reproduction would also be useful and this mechanism must also be related to dissipation. It is also important to explain how and why these molecules would have transformed into the information carrying molecules that they are today. 

A possible mechanism for replication of RNA and DNA without the need for enzymes (and thus information content and reproductive fidelity) called Ultraviolet and Temperature Assisted Reproduction (UVTAR) has been suggested (Michaelian, 2009b, 2011b) which employs the physical characteristics present on Archean Earth's surface; high surface temperature $\approx$ 80$^\circ$C (Lowe and Tice, 2004), intense UV light, and moderate $\approx$ 5$^\circ$C day/night temperature cycling of the ocean surface (Michaelian, 2009a, 2012). It is probable that nucleotides would have proliferated on the ocean surface due to their propensity to act as catalysts for UV dissipation (see section \ref{sec:autocatalysis}) and due to their resistance to UV destruction, and that some phosphate bonding among these would have occurred under the action of UV light (Strigunkova et al., 1986).

As the Earth's surface temperature cooled to just below the denaturing temperature of certain RNA or DNA segments, then during the day absorption of infrared and visible light on water, and absorption of UV on nucleotides, would be sufficient to raise the local surface water temperature to above the denaturing temperature of RNA or DNA, separating the double strands. During the cool periods overnight, these single strands could have acted as templates for the production of new double strand. This proposed mechanism bears resemblance to polymerase chain reaction (PCR) which is today used routinely in the laboratory to amplify a particular segment of DNA or RNA (Mullis, 1990). 

It is probable that as the Earth's surface cooled further, the daytime increase in surface temperature was insufficient to permit denaturation of RNA or DNA by itself and so those strands which happened to code for an amino acid which could act as a simple denaturing enzyme would become more prevalent (Michaelian, 2011b). These amino acids could have been those with an aromatic ring; tyrosine, tryptophan, phenylalanine, or histidine, which would have acted as antenna molecules to attract more UV light for more local heating. Information content and replication fidelity leading to evolution through natural selection would therefore have arisen as a thermodynamic response to retaining entropy production through the multiplication of these UV dissipating pigments as the seas cooled further.

The details of the mechanism have been presented elsewhere (Michaelian, 2009b, 2011b), suffice to say here that this proposed mechanism enjoys a number of advantages over previous theories for the origin of life. First and foremost, life's origin is clearly recognized as a dissipative process, and, furthermore, it is tied directly to the dissipation of the most intense free energy source available at the Earth's surface. Second, since the mechanism of replication relies on the physical characteristics of the Archean environment, and not on specialized enzymes, there is no requirement on initial reproduction fidelity to protect information content of RNA or DNA. Under the UVTAR mechanism, information content was not required for replication. Third, since there would be a slight denaturation advantage to right handed RNA or DNA during the late afternoon when the ocean surface temperature is highest and the submarine light at the surface is slightly right-circularly polarized, the present theory could explain the chirality of life (Michaelian, 2010).

\subsection{Dissipation through organic molecules throughout the universe}
\label{sec:lifeuniverse}

On Earth, organic molecules are found only in association with water. As described above, this is most likely related to the efficiency of organic pigments dissipating solar photons using the high frequency water vibrational modes to facilitate their de-excitation. Without water they are poor photon dissipaters and easily destroyed by photochemical reactions. This is probably the primordial reason for the association of life with water.

Organic molecules have been found in significant quantities on the surfaces of inner orbit comets. It is probable that these molecules were not formed in the atmospheres of red giant stars and later collected by these comets as suggested by Hoyle (1978), but rather that they formed on the comets themselves as the comet passed sufficiently close to the Sun to cause regions of liquid water to form on the surface of the comet. As on Earth, on the comet surface, the proliferation of these molecules to beyond their expected equilibrium concentrations would be a thermodynamic imperative, a result of their catalytic activity in dissipating high energy photons from the Sun.

However, it is entirely possible that a solvent other than water, with high frequency vibrational modes that can couple to the vibrational modes of the same or similar UV absorbing pigment might exist on another planet. In this case, the temperature range for analogous dissipating ``life" in our universe might not be confined to that required for liquid water. This would greatly increase the variety of possible life and the expectation of finding evolved dissipating life spread on the surface of another planet.

However, the evolution of our particular form of life based RNA and DNA and water, assuming the correctness of the above proposed mechanism for its origin, would be very dependent on the particular initial conditions of the planet. For example, an intense UV photon flux, and liquid water temperatures that gradually dropped below the denaturing temperature of RNA and DNA, could only happen on a planet at a similar distance from a similar star as our Sun.

Finally, this delicate dependence of Earth biological evolution on its initial conditions, and thus probably also on external perturbations, suggests that it was probably not inevitable that life as we know it appeared on Earth. For example, the elastic dispersion into a greater volume of an initially collimated photon beam also produces a significant amount of entropy, contributing, in fact, almost one half of the entropy production on Venus (Michaelian, 2012). It may be that during ice-ages, the majority of the entropy production was shifted from photon dissipation by photosynthetic organisms to photon elastic dispersion from ice, or, more probably, a combination of the two forms of dissipation which becomes competitive with photon dissipation by organic pigments alone under the prevailing conditions.

\section{Conclusions}
\label{sec:conclusions}

As for all irreversible processes, life is directly dependent on the dissipation of a generalized thermodynamic potential. This potential is today, and always has been, the high energy solar photon flux, and this dissipation is the largest source of entropy production on Earth. There is evidence that life has co-evolved with its abiotic environment, adjusting the physical characteristics of Earth such that the most intense part of the solar photon spectrum can arrive at the Earth's surface where it can be dissipated by life.

About 71\% of the entropy production due to the dissipation of solar photons occurs at Earth's surface. Earth produces  approximately 70\% more entropy per unit surface area than either of its neighbors (Michaelian, 2012). Both of these facts are probably due to life on the surface of Earth.

Classical irreversible thermodynamics indicates that if organic pigments act to catalyze the dissipation of the solar photon potential then their quantity on Earth's surface can be expected to exceed greatly the expected equilibrium values and the entropy production of the dissipation process will increase accordingly. This is undoubtedly the reason for the proliferation of organic pigments over the Earth's surface. For a system with many more coupled irreversible processes operating than considered here, this thermodynamic result may explain the motor behind all biological and biotic-abiotic evolution. From this thermodynamic dissipative perspective, animals would only have thermodynamic relevance in how they assist in the growth and spread of plants over land and cyanobacteria over the water. 

Structuring due to dissipation is a universal phenomena. Forms of dissipation are varied. Life as we know it is only one based on the direct dissipation of visible or UV light. Organic pigments are the most efficient and robust dissipater in this wavelength region given Earth's physical and chemical characteristics. On the solar systems of different stars it may be that very different dissipaters exist, some based on different organic molecules dissipating different photon chemical potentials (spectra), others based on a different type of solvent molecule that couples to the vibrational modes of the excited organic molecule, and this solvent could be operating in different physical (temperature, pressure, etc.) regimes. It is well known, for example, that organic molecules in the clouds of Venus are absorbing UV and visible light and dissipating this energy into the heat that drives the great southern vortex (ESA, 2006). It would be interesting to show that these dissipating molecules have proliferated to beyond their expected equilibrium concentrations and thus we must acknowledge a different type of living ecosystem operating on Venus. 

\ack
The financial assistance of DGAPA-UNAM, grant numbers IN112809 and IN103113 is greatly appreciated.

\section*{References}
\begin{thereferences}

\item \'AHearn M F et al. 2005 Deep impact: excavating comet Tempel 1  {\em Science} {\bf 310} 258  DOI: 10.1126/science.1118923

\item Callahana M P et al. 2011 Carbonaceous meteorites contain a wide range of extraterrestrial nucleobases, {\bf PNAS} Early Edition, 13995

\item  Cnossen I, Sanz-Forcada J, Favata F, Witasse O, Zegers
T, and Arnold N F 2007 The habitat of early life: Solar X-ray and UV
radiation at Earth's surface 4--3.5\,billion years ago {\em J. Geophys. Res.}
{\bf 112} E02008 http://dx.doi.org/10.1029/2006JE002784 {doi:10.1029/2006JE002784}

\item Conde F R, Churio M S, Previtali C M 2000 The photoprotector mechanism of mycosporine-like amino acids. Excited-state properties and photostability of porphyra-334 in aqueous solution {\em Journal of Photochemistry and Photobiology B: Biology }{\bf56} 139–144

\item de Donder T 1936 Thermodynamic Theory of Affinity: A Book of Principles. Oxford, England: Oxford University Press

\item European Space Agency 2006, June 27 Double vortex at venus south pole unveiled  {\em ScienceDaily} Retrieved January 10, 2012, from http://www.sciencedaily.com­ /releases/2006/06/060627104232.htm

\item Gadgil C J and Kulkarni D B 2009 Autocatalysis in Biological Systems {\em AIChE Journal} {\bf 55} No. 3, 556--562 

\item  Gates D M 1980 Biophysical Ecology ISBN~0-387-90414-X,
Springer-Verlag, New York

\item Herrmann F and W\"urfel P 2005 Light with nonzero chemical potential {\em Am. J. Phys.} {\bf 73} 717--721

\item Hoyle F and Wickramasinghe N C 1978 Lifecloud -- The Origin of
Life in the Universe, ISBN~0-460-04335-8, J.~M.~Dent and Sons, London

\item Jones I, George G, Reynolds C 2005 Quantifying effects
of phytoplankton on the heat budgets of two large limnetic enclosures {\em Freshwater Biology} {\bf 50} 12391247

\item Kleidon A 2008 Entropy Production by Evapotranspiration
and its Geographic Variation {\em Soil \& Water Res.} {\bf 3} S89–S94

\item Lovelock J E 1988 The Ages of Gaia: A Biography of
Our Living Earth W. W. Norton \& Company, New York

\item Lowe D R and Tice M M 2004 Geologic evidence for Archean atmospheric and climatic evolution: Fluctuating levels of CO2, CH4, and O2 with an overriding tectonic control {\em Geology} {\bf 32} 493-496 

\item Michaelian K 2009a Thermodynamic function of life (PDF) arXiv. http://arxiv.org/abs/0907.0040

\item Michaelian K 2009b Thermodynamic origin of life (PDF) arXiv. http://arxiv.org/abs/0907.0042

\item Michaelian K 2010 Homochirality through photon-induced melting of RNA/DNA: the thermodynamic dissipation theory of the origin of life. Available from {\em Nature Precedings} http://hdl.handle.net/10101/npre.2010.5177.1

\item Michaelian K and Manuel O 2011a Origin and evolution of life constraints on the solar model {\em Journal of Modern Physics} {\bf 2} No. 6A, 587-594 DOI: 10.4236/jmp.2011.226068

\item Michaelian K 2011b Thermodynamic dissipation theory for the origin of life {\em Earth Syst. Dynam.} {\bf 2} 37–51 doi:10.5194/esd-2-37-2011

\item Michaelian K 2012 Biological catalysis of the hydrological cycle: life's thermodynamic function {\em Hydrol. Earth Syst. Sci.} {\bf 16} 2629-2645, 2012.
www.hydrol-earth-syst-sci.net/16/2629/2012/
doi:10.5194/hess-16-2629-2012

\item  Miller S L and Urey H C 1959 Organic compound synthesis on the
primitive earth {\em Science} {\bf 130} 245--251

\item Mullis K 1990 The unusual origin of the Polymerase Chain Reaction {\em Scientific American} April, 56–65

\item  Onsager L 1931 Reciprocal Relations in Irreversible Processes, I.
{\em Phys. Rev.} {\bf 37} 405--426

\item Pecourt JM L, Peon J, and Kohler B 2000 Ultrafast internal conversion of electronically excited RNA and DNA nucleosides in water {\em J. Am. Chem. Soc.} {\bf 122} 9348-9349

\item Pisani D,  Poling L L, Lyons-Weiler M and Hedges S B 2004 The colonization of land by animals: molecular phylogeny and divergence times among arthropods {\em BMC Biology} {\bf 2:1} doi:10.1186/1741-7007-2-1
 
\item  Prigogine I 1967 Thermodynamics of Irreversible Processes, Wiley,
New York

\item Ribas I, Guinan E F, Güdel M and Audard M 2005 Evolution of the solar activity over time and effects on planetary atmospheres. I. High-energy irradiances (1 to 1700 Å) {\em ApJ.} {\bf 622} 680-- 694

\item  Sagan C 1973 Ultraviolet selection pressure on the earliest
organisms {\em J. Theor. Biol.} {\bf 39} 195--200

\item  Sagan C and Chyba C 1997 The early faint Sun paradox: organic
shielding of ultraviolet-labile greenhouse gases {\em Science} {\bf 276} 1217--1221

\item Serrano-Andr\'es L and Merch\'an M 2009 Are the five natural DNA/RNA base monomers a good choice from natural selection? A photochemical perspective, {\em Journal of Photochemistry and Photobiology C: Photochemistry Reviews} {\bf 10} 21–32

\item Sinha R P, Sinha  J P, Groniger A, Hader D-P 2002 Polychromatic action spectrum for the induction of a mycosporine-like
amino acid in a rice-field cyanobacterium, Anabaena sp. {\em Journal of Photochemistry and Photobiology B: Biology} {\bf 66} 47–53

\item Strigunkova TF, Lavrentiev G A, and Otroshchenko V A 1986 Abiogenic synthesis of oligonucleotides on kaolinite under the
action of ultraviolet radiation {\em J Mol Evol} {\bf 23} 290--293

\item Wu W, Liu Y, Wen G 2011 Spectral solar irradiance and its entropic effect on Earth’s climate {\em Earth Syst. Dynam. Discuss.} {\bf 2} 45–70 www.earth-syst-dynam-discuss.net/2/45/2011/ doi:10.5194/esdd-2-45-2011.

\end{thereferences}

\end{document}